\documentclass{aa} 
\usepackage{graphicx,amssymb,txfonts}

\begin{document}
\title{Multiple accretion events as a trigger for Sgr A* activity}
\author{Bo\.zena Czerny\inst{1}
         \and Devaky Kunneriath\inst{2}
         \and  Vladim\'{\i}r Karas\inst{2}
         \and Tapas K. Das\inst{3}}
\institute{Nicolaus Copernicus Astronomical Centre, Bartycka 18, P-00716 Warsaw, Poland
	\and Astronomical Institute, Academy of Sciences, Bo\v{c}n\'{\i} II 1401, CZ-14100 Prague, Czech Republic
	\and Harish Chandra Research Institute, Allahabad 211019, India}
\abstract{Gas clouds are present in the Galactic centre, where they orbit around the supermassive black hole. Collisions between these clumps reduce their angular momentum, and as a result some of the clumps are set on a plunging trajectory. Constraints can be imposed on the nature of past accretion events based on the currently observed X-ray reflection from the molecular clouds surrounding the Galactic centre.}{We discuss accretion of clouds in the context of enhanced activity of Sagittarius~A* during the past few hundred years. We put forward a scenario according to which gas clouds bring material close to the horizon of the black hole on $\lesssim0.1$ parsec scale.}{We have modelled the source intrinsic luminosity assuming that multiple events occur at various moments in time. These events are characterized by the amount of accreted material and the distribution of angular momentum. We parameterized the activity in the form of a sequence of discrete events, followed the viscous evolution, and calculated the luminosity of the system from the time-dependent accretion rate across the inner boundary.}{Accreting clumps settle near a circularization radius, spread there during the viscous time, and subsequently feed the black hole over a certain period. A significant enhancement (by factor of ten) of the luminosity is only expected if the viscous timescale of the inflow is very short. On the other hand, the increase in source activity is expected to be much less prominent if the latter timescale is longer and a considerable fraction of the material does not reach the centre.}{A solution is obtained under two additional assumptions: (i) the radiative efficiency is a decreasing function of the Eddington ratio; (ii) the viscous decay of the luminosity proceeds somewhat faster than the canonical $L(t){\propto}t^{-5/3}$ profile. We applied our scheme to the case of G2 cloud in the Galactic centre to obtain constraints on the core-less gaseous cloud model.}
\date{Received 19 September 2011; Accepted 16 May 2013}
\authorrunning{B. Czerny, D. Kunneriath, V. Karas, \& T. K. Das}
\titlerunning{Multiple accretion events for Sgr A* activity}
\maketitle

\section{Introduction}
\label{sec:introduction}
The centre of the Milky Way galaxy contains a supermassive black hole of mass $M_{\rm BH}=m_{\bullet}\,4.4 \times 10^6M_{\odot}$ (where the dimensionless $m_{\bullet}$ is about unity; Genzel et al. 2010). The present activity of the Sagittarius~A* (Sgr A*) nuclear source is very low, probably because the accretion rate is very low and radiatively inefficient (Eckart et al. 2005; Melia 2007). The mass accretion rate is estimated within the range from $\simeq 10^{-9} M_{\odot}$ yr$^{-1}$ to a few times $10^{-7} M_{\odot}$ yr$^{-1}$, as implied by the measurement of the Faraday rotation (Marrone et al. 2007; Ferri\`ere 2009). 

Because many stars are present in the region and form a dense nuclear cluster, the material to support the observed level of activity can be provided by stellar mass loss in its entirety (Wardle \& Yusef-Zadeh 1992; Melia 1992; Coker \& Melia 1997; Loeb 2004; Rockefeller et al. 2004; Cuadra et al. 2005, 2006; Moscibrodzka et al. 2006). In fact, a number of papers address the question of why only a fraction of the material from stellar winds reaches the black hole (e.g., Quataert 2004; Shcherbakov \& Baganoff 2010).

Despite the low level of current activity of the Galactic centre, Integral and XMM-Newton observations of the X-ray reflection from molecular clouds in the Sgr A* region seem to imply that just a few hundred years ago \object{Sgr~A*} was orders of magnitude brighter than it is currently (Sunyaev et al. 1993, 1998; Koyama et al. 1996; Muno et al. 2007; Inui et al. 2009; Ponti et al. 2010, 2011; Terrier et al. 2010; Nobukawa et al. 2011; Capelli et al. 2012; Ryu et al. 2013). As shown by Cuadra et al. (2008), stellar winds do not explain such an enormous and relatively quick change in luminosity. Instead, an occasional inflow of clumps of the fresh material is more likely, and one such cloud (G2) is now approaching the Galactic centre (Gillessen et al. 2012, 2013). 

It is thought that the G2 cloud should reach the closest distance of $\sim2200 R_{\rm S}$ in mid 2013 or early 2014.\footnote{Length units of Schwarzschild radius are used as a natural scale of the problem, $R_{\rm S}=2GM_{\rm BH}/c^2\sim1.3\times10^{12}m_{\bullet}$~[cm], where $G$ denotes gravitational constant, $c$ is the speed of light.} Then the cloud should become disrupted by its interaction with the ambient medium. The material will then gradually diffuse towards the black hole. It has been suggested that the silhouette of the supermassive black hole will be smeared as a result of rising luminosity of Sgr~A* (Moscibrodzka et al. 2012).

In this paper we focus our attention on a specific subtopic, namely, we study the constraints on the number, frequency and strength of such events in the past, based on the currently observed X-ray reflection from molecular clouds surrounding Sgr~A*. This question is also interesting in the context of the variability that the source exhibits on different timescales in different energy bands, but currently it is difficult to understand whether the vastly disparate scales are driven by the same underlying mechanism (Witzel et al. 2012). At this stage we approach the problem by constructing a simplified scheme, and we find that a relatively simple set-up leads us to very reasonable estimates and dependencies, although we cannot yet develop a precise model. We demonstrate that repetitive accretion events are able to explain the basic properties of the changing activity of Sgr~A* black hole during the past several hundred years. A simple scheme allows us to set interesting constraints on the model parameters. 

In the next section we describe the model and illustrate its application in two versions -- for a single accretion event and for a sequence of repeated events. Then, in sect. \ref{sec:results}, we use this scheme to put constraints on a possible form of the observed lightcurve of the signal reflected from clouds surrounding Sgr~A* at a certain distance. In particular, we concentrate our attention on the Sgr~B2 and Sgr~C1 Galactic centre clouds for which sufficiently accurate (3D) positions have been reported in the literature. We discuss different aspects of the model in sect. \ref{sec:discussion}, and we summarize our conclusions in sect. \ref{sec:conclusions}. 

At present we cannot proceed far beyond the qualitative exposition of the model and the discussion of basic quantitative constraints, but this should be possible in future when more data points are added to the observed lightcurve and the positions of the reflecting clouds are determined with better accuracy for more clouds in the region.

\section{Method}
We model the intrinsic luminosity of Sgr~A* and the luminosity of the reflection from a distant molecular cloud assuming discrete (multiple) accretion events that happen at various moments in time and are characterized by a distribution of angular momentum of the infalling clouds. We develop the model in two steps: first, in the next section we describe the relevant aspects of the individual accretion events when the discrete clouds get close to the centre and then their material becomes dispersed and captured by the black hole. Then we proceed to the case of multiple accretion events that recur over a period of time.

\subsection{Description of a single accretion event}
Direct accretion of a gaseous clump is not possible if the angular momentum of the infalling gas exceeds a certain threshold. If the inflowing clump (or a stellar body) becomes tidally disrupted before reaching the pericentre, part of its material is lost, and the remaining fraction gathers to form a ring at the circularization radius. Later, as the disc viscosity starts to operate the ring can disperse in radius and trigger an accretion event. This phenomenon was studied analytically and numerically in a number of papers, especially in the context of putative stellar disruptions and the resulting flares occurring close to a massive black hole (e.g. Frank \& Rees 1976; Evans \& Kochanek 1989; MacLeod et al. 2012; Reis et al. 2012, and further references cited therein). In the Galactic centre the flares are known to occur on a daily basis and can be detected over a broad range of wavelengths (Eckart et al. 2008, 2012; Kunneriath et al. 2010).

Here we aim at studying a broad range of parameters with a simplified scheme, so we follow the analytical approach to the description of the cloud disruption and the subsequent decay phase of the X-ray lightcurve. We start from equations governing the evolution of an accretion disc (Lynden-Bell \& Pringle 1974),
\begin{equation}
{\partial \Sigma \over \partial t} = {1 \over 2 \pi R }{\partial \dot M \over \partial R},
\label{eq:continuity}
\end{equation} 
\begin{equation}
\dot M = 6 \pi R^{1/2} {\partial \over \partial R}\left(R^{1/2} \nu \Sigma\right),
\label{eq:angmom}
\end{equation}
which operate on the viscous timescale; $\Sigma (R)$ and $\dot M (R)$ are the disc surface density and the accretion rate at a given radius $R$. 

Kinematic viscosity $\nu$ can be set as either constant or a power-law function of the radius,
\begin{equation}
\nu = \nu_0 \left({R \over R_0}\right)^n.
\end{equation}
Such a prescription is more general than the standard Shakura--Sunyaev scheme. By adopting specific values of the power-law index $n$, one can mimic the gas-dominated regime, the radiation-pressure dominated regime, or the isothermal disc regime of the standard model (see Kato et al. 1998; Zdziarski et al. 2009).  

These equations need to be supplemented with the time-dependent description of the disc heating and cooling processes, and solved numerically. Analytical solutions are possible under certain simplifying constraints. What is important is that the analytical solutions for time and radial dependences of the disc surface density and the local accretion rate are the Green function of the problem. A general solution can thus be obtained as a superposition of these elementary solutions. 

Green's functions of the problem adopt the following form:
\begin{equation}
G_{\Sigma}(R,t) = {2 \Sigma_0 |\mu| \xi ^{1/\mu - 9/2} \over \tau}\,\exp\left[- {2 \mu^2 (\xi^{1/\mu} + 1)\over \tau}\right]\, I(\mu;\xi,\tau),
\label{eq:sigma}
\end{equation}
\begin{equation} 
G_{\dot M}(R,t) = {\dot M_0 |\mu| \over \tau}\, {\partial \over \partial \xi} \xi^{1/2}\,\exp\left[- {2 \mu^2 (\xi^{1/\mu} + 1) \over \tau}\right]\, I(\mu;\xi,\tau),
\label{eq:mdot}
\end{equation}
where 
\begin{equation}
 I(\mu;\xi,\tau) \equiv I_{|\mu|}\left[{4 \mu^2 \xi^{1/(2\mu)} \over \tau}\right]
\end{equation}
denotes the modified Bessel function of the first kind,
\begin{equation}
\dot M_0 = {4 \pi R_0^2 \Sigma_0 \over t_{\rm visc}(R_0) }, ~~ \mu = {1 \over 4 - 2n}, ~~ \tau= {t \over t_{\rm visc}(R_0)}, ~~ t_{\rm visc} = {2 R_0^2 \over 3 \nu},
\end{equation}
and $\xi=R/R_0$ is the dimensionless radius (scaled with the outer radius where the mass is supplied). Equation (\ref{eq:sigma}) was studied by, e.g., Mineshige \& Wood (1989), Lyubarskii (1997), and Kotov et al. (2001). Equation (\ref{eq:mdot}) was originally derived and discussed by Zdziarski et al. (2009). An implicit assumption underlying these equations is that the initial distribution of the mass adopts the form of an equatorial ring encircling the black hole at constant radius, $R=R_0$, with the initial profile of the surface density in the form $\Sigma(R)_{|t=0} = \Sigma_0\,\delta(R- R_0)$, where $M_0=2\pi R_0\Sigma_0$ is the total mass of the ring.

We are interested in the temporal evolution of luminosity. Most energy is liberated in the form of radiation at the inner part of the accretion disc; therefore, in further considerations, we can use the dimensionless asymptotic expression (which is strictly valid only at $\xi = 0$; Zdziarski et al. 2009),
\begin{equation}
\tilde G_{\dot M}(\tau) = {\dot M_0  (2 \mu^2)^{\mu} \over \Gamma(\mu) } \,\tau^{-1 - \mu}\,\exp\left(-{ 2 \mu^2\over \tau}\right),
\end{equation}
where $\Gamma$ is the Euler gamma function. 

The duration of the bright phase of such an event is expected to be close to the viscous time at the circularization radius $R_0$, with a sharp rise and  the decay phase in the form of  a power law with a slope equal to $-(1+\mu)$. Thus, the duration of the brightest phase of the large angular momentum event would be close to the viscous timescale of the standard disc,
\begin{equation}
t_{\rm visc} = 1.8\;m_{\bullet}^{-1/2} \left({R_{\rm 0} \over 2200R_{\rm S}}\right)^{3/2}\left({R_0 \over H(R_0)}\right)^2 \left({0.1 \over \alpha}\right)~~~[\mbox{yr}], 
\end{equation}
strongly depending on the value of the circularization radius $R=R_0$ and the ratio of the disc thickness to the radius, $H(R)/R$. 

In the case of a single discrete event, a dimensional formula for the time-dependent accretion rate of material overflowing across the inner edge is most conveniently expressed as 
\begin{equation}
\dot M(t) = {M_{\rm 0} \over  t_{\rm visc}} {(2 \mu^2)^\mu \over \Gamma(\mu)}\, \tau^{-1 -\mu}\, \exp\left(-{2 \mu^2 \over \tau}\right),
\label{eq:mdot1}
\end{equation}
where $M_{\rm 0}$ is the mass deposited in the ring, and $\mu$ the dimensionless parameter of the flow properties. The likely values of $\mu$ cover an interval $\langle1/3,1\rangle$, depending on the interplay between viscous and cooling processes. 

\subsection{Sequence of multiple accretion events}
The set of equations (\ref{eq:continuity})--(\ref{eq:angmom}) is linear in $\dot M$ and $\Sigma$, so the elementary solutions can be superposed. The solution for a specific single event is a Green function of the problem, whereas a general case can be then expressed as an integral (Zdziarski et al. 2009). We are interested in a set of $N$ discrete events, so we express the resulting $\dot M(t)$ as a sum,
\begin{equation}
\dot M(t) = \sum\limits_{i=1}^N {M_{0,i} \over  t_{{\rm visc,}i}} {(2 \mu^2)^\mu \over \Gamma(\mu)}\, \tau_i^{-1 -\mu}\, \exp\left(-{2 \mu^2 \over \tau_i}\right),
\end{equation}
where the mass  and the radius of injection (and so the angular momentum) can be different for every event (denoted by index~$i$). 

Naturally, a contribution of the given event to the total lightcurve  starts at the moment of injection, $t=t_i$, i.e.,
\begin{equation}
\tau_i = {t - t_i \over t_{\rm visc}(R_i)},
\end{equation}
and the contribution of a term $i$  is zero for $t < t_i$. 

Although the standard accretion disc is not present in Sgr~A*, episodic accretion events of individual falling clouds are possible and even likely. If the time separation of events is comparable to or shorter than the decay timescale, the events can overlap, which may lead to a complicated time profile of the resulting total luminosity. 

\subsection{Bolometric and X-ray luminosity of accretion flow}
A significant fraction of radiation from accretion is released close to the black hole, i.e., in the deep potential well. We thus assume that the bolometric luminosity $L_{\rm bol}$ of the flow is given by the accretion rate at the inner edge of the flow. Radiation is produced with the accretion efficiency $\eta$,
\begin{equation}
L_{\rm bol} = \eta \, \dot M(t) \, c^2,
\end{equation}
where $\dot{M}(t)$ is the time-dependent accretion rate.  

In general, the accretion efficiency depends on the black hole spin, as well as on the accretion rate. Efficiency is higher for fast-rotating black holes accreting at a fraction of Eddington rate mainly because the innermost stable circular orbit (ISCO) gets closer to the horizon, and the inferred radiation efficiency grows as the spin increases. The efficiency drops down at low Eddington ratios when the flow becomes optically thin and radiatively inefficient. However, the spin of Sgr A* supermassive black hole is not well constrained (e.g. Broderick et al. 2011). The flow efficiency also depends, rather sensitively, on the fraction of heat that goes directly to electrons, and this factor is usually taken as a free parameter of the model. 

The efficiency is expected to be low for sources that are well below the Eddington luminosity (see Fig.~4 of Narayan \& McClintock 2008). For Sgr~A* the mean accretion rate is estimated as $10^{-8} M_{\odot}$ yr$^{-1}$, i.e. a very low value, and the average broad-band luminosity is about $10^{36}$ erg~s$^{-1}$. This gives the efficiency parameter about $\eta\simeq10^{-3}$.

In the present paper we consider two options: either we fix the efficiency at a constant value $\eta\simeq10^{-3}$, or we adopt a more general luminosity-dependent trend that has been proposed in the context of starving black holes (Hopkins et al. 2006; Narayan \& McClintock 2008),
\begin{equation}
\log \eta = 1 + \log \left({\dot M \over \dot M_{\rm Edd}}\right),
\label{eq:eta}
\end{equation}
where
\begin{equation}
\dot M_{\rm Edd} ={ L_{\rm Edd} \over 0.1 c^2}.
\end{equation}
The latter factor is equal to $5.7 \times 10^{24}$ g s$^{-1}$ for the Sgr~A* black hole mass. 

\begin{figure}
\begin{flushright}
\includegraphics[angle=0,width=0.48\textwidth]{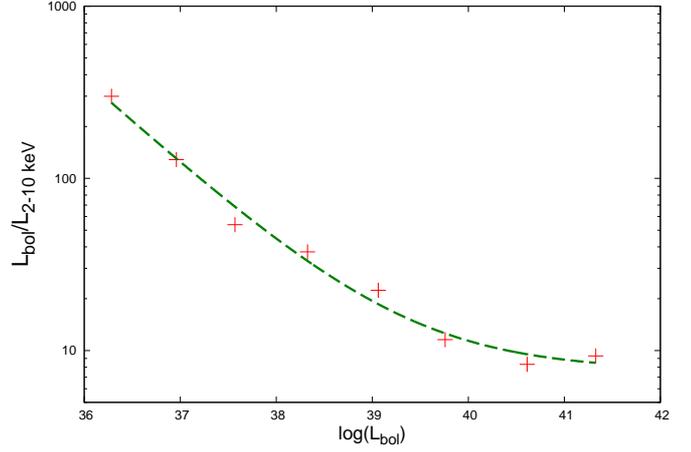}
\end{flushright}
\caption{The ratio of the bolometric luminosity to the 2--10 keV X-ray luminosity according to the theoretical modelling (Moscibrodzka et al. 2012) for Sgr~A$^*$ supermassive black hole.}
\label{fig:bolo_cor}
\end{figure}

The observational constraints for the past activity refer to the X-ray luminosity. Therefore, we also need a conversion from the bolometric flux, $L_{\rm bol}$, to 2--10 keV X-ray flux, $L_{2\mbox{--}10\, \rm keV}$. We employ two estimates: either we set the conversion factor $\eta_X$ to the value of 10, or we use a more realistic luminosity-dependent relation. We derive the latter relation on the basis of the accretion flow models of Moscibrodzka et al. (2012). In Fig.~\ref{fig:bolo_cor} the points show the values derived from these theoretical (numerically solved) models, whereas the continuous line is the best fit function
\begin{equation}
\eta_X = \left({1.32 \times 10^{41} \over L_{\rm bol}}\right)^{0.5} + 7.86,
\label{eq:bolo_cor}
\end{equation}
which we use to describe the luminosity-dependent trend in our code.

\subsection{Reprocessing by an extended cloud}
Molecular clouds have an extended size. In particular, we consider reprocessing of the light signal by the prominent Sagittarius B2 cloud. Modelling the cloud reflection has to take the effect of the cloud's non-negligible dimension into account. To this end we follow the parameterization by Odaka et al. (2011), who discuss different possible positions of the B2 cloud (see their Fig.~3). The recent analysis by Ryu et al. (2009, 2013) indicates that the B2 cloud is situated at approximately a right angle with respect to the line of sight between Sgr A* and the observer. 

We assume the cloud to be spherically symmetric and optically thin. (We neglect the role of multiple scattering.) Despite the substantial size of the cloud, it is still small in comparison with the distance between the cloud and Sgr A*. This allows us to simplify computations by neglecting the curvature of surfaces of constant timedelay. 

We consider three examples of the density distribution within the cloud.  In the first case we set density $\rho_{\rm o}={\rm const}$. The lightcurve of the reprocessed radiation in K$\alpha$ line can be then expressed as
\begin{equation}
L_{\rm rep}(t) = \eta_{\rm line} \left({R \over 2 D_{\rm o}}\right)^2\! \kappa\, \rho_{\rm o}\, R\! \int_{-1}^{1} \left(1 - x^2\right)\, L_{\rm bol}\big(t - \Delta t(x)\big)\,dx,
\end{equation}
and
\begin{equation}
\Delta t (x) = {\sqrt 2 R x - D_{\rm o}\over c}\,,
\end{equation}
where $R$ is the cloud radius, $x$ dimensionless coordinate within the cloud, $\kappa$ the opacity,  $L_{\rm bol}$ the time-dependent bolometric luminosity of Sgr A*, $\eta_{\rm line}$ the efficiency of converting bolometric luminosity into iron K$\alpha$ line flux, and $D_{\rm o}$ the distance between Sgr~A* and the centre of the cloud. 

Sunyaev \& Churazov (1998) derive the efficiency of the 6.4 keV iron line formation to be $\eta_{\rm line}\simeq0.10$ of the monochromatic luminosity at 8 keV (assuming the solar abundance of iron). Ryu et al. (2013) used this coefficient in combination with the assumption of the photon index of 1.6, which allowed these authors to find a connection between the 6.4 keV line flux and 2--10 keV X-ray continuum flux, namely, $L_{2\mbox{--}10\, \rm keV} = 1.3 L_8$. We followed this relation and calculate $\eta_{\rm line}$ using eq.~(\ref{eq:bolo_cor}). Since the reprocessing took place in the past when the source was bright, and because the relation between the present bolometric luminosity and the 2--10 keV flux is close to 10 and does not vary strongly, the value of $\eta_{\rm line}$ comes out close to 0.0077. This value varies only weakly with the X-ray luminosity.

Next, we considered a two-component structure of the cloud (core plus envelope) in the form of a central spherical nucleus  and a surrounding atmosphere. The simplest approximation of this structure assumes two different (constant) values of density within the core and the envelope. Finally, as the most realistic representation we adopted a gradually decreasing density profile in the envelope, which encloses the dense nucleus. In the latter case we were able to obtain an analytical description only in special cases of the density profile, namely,
\begin{eqnarray}
n(R) & = & n_{\rm c}, ~~~~~~ R < R_{\rm c}  \nonumber \\
     & = & n_{\rm c} \left({R \over R_{\rm c}}\right)^{-1}, ~~~~~~  R_{\rm c}<R< R_{\rm env}
\label{eq:nr}
\end{eqnarray}
where $n_{\rm c}$ and $R_{\rm c}$ are the density and the radius of the inner core of the cloud, and $R_{\rm env}$ is the total size of the cloud envelope. The slope $-1$ in eq. (\ref{eq:nr}) characterizes the decreasing density of the cloud envelope. In this case only one integral on the constant timedelay surface can be obtained analytically, whereas the second one has to be evaluated numerically. Instead of the factor $(1 - x^2)$ we obtain an integral expression,
\begin{equation}
I(x) = \int_{-\sqrt{ 1 - x^2}}^{\sqrt{ 1 - x^2}} I(x,y)\,dy,
\end{equation}
where $I(x,y)$ is given by the following expressions:
\begin{eqnarray}
I(x,y)& =& \xi_{\rm o} \ln {1 + A \over \xi_{\rm o} +B } + \nonumber \\ 
   &&  \xi_{\rm o} \ln {\xi_{\rm o} - B \over 1 - A} + 2 B , ~~ {\rm for} ~~  y^2 < \xi_{\rm o}^2 - x^2, \\
I(x,y)& = & \xi_{\rm o} \ln {1 + A \over 1 - A}, ~~~~ {\rm for}  ~~ y^2 > \xi_{\rm o}^2 - x^2,
\end{eqnarray} 
with $\xi_{\rm o} = R_{\rm c}/R_{\rm env}$, $A = (1 - x^2 - y^2)^{1/2}$ and $B = (\xi_{\rm o}^2 - x^2 - y^2)^{1/2}$.

In the more general case of an arbitrary slope of the radial density profile of the cloud envelope, the whole integral across the surface needs to be evaluated numerically. We ignore this complication because the three above-mentioned examples are enough to capture the effect and describe the extreme cases of the density profile. Also, we neglect the role of multiple scattering (included in the computations of Odaka et al.\ 2011) because we aim at a simple and robust estimation of the duration of reprocessing instead of any detailed computation of the spectral shape of the reflected component.

\section{Results}
\label{sec:results}
We have modelled the time evolution of the clumpy inflow episodes onto the black hole with the goal of understanding the observed signal from Sgr A*, namely, its past variations and the present level of activity. Naturally, we would also like to know what constraints can be inferred for the future behavior of the source.  We start with a single accretion event, then we discuss multiple events and the reprocessing of the intrinsic radiation by molecular clouds surrounding the object.

\subsection{Case of single accretion event}
We consider two examples as the basis of the subsequent discussion. They are also of immediate observational interest as the expected evolution of the G2 cloud after the predicted disruption (Gillessen et al. 2012, 2013) and as a significant event that could have been responsible for activity in the past (Ponti et al. 2010, 2011; Miralda-Escud\'e 2012).

\begin{figure*}
\begin{flushright}
\includegraphics[angle=0,width=0.49\textwidth]{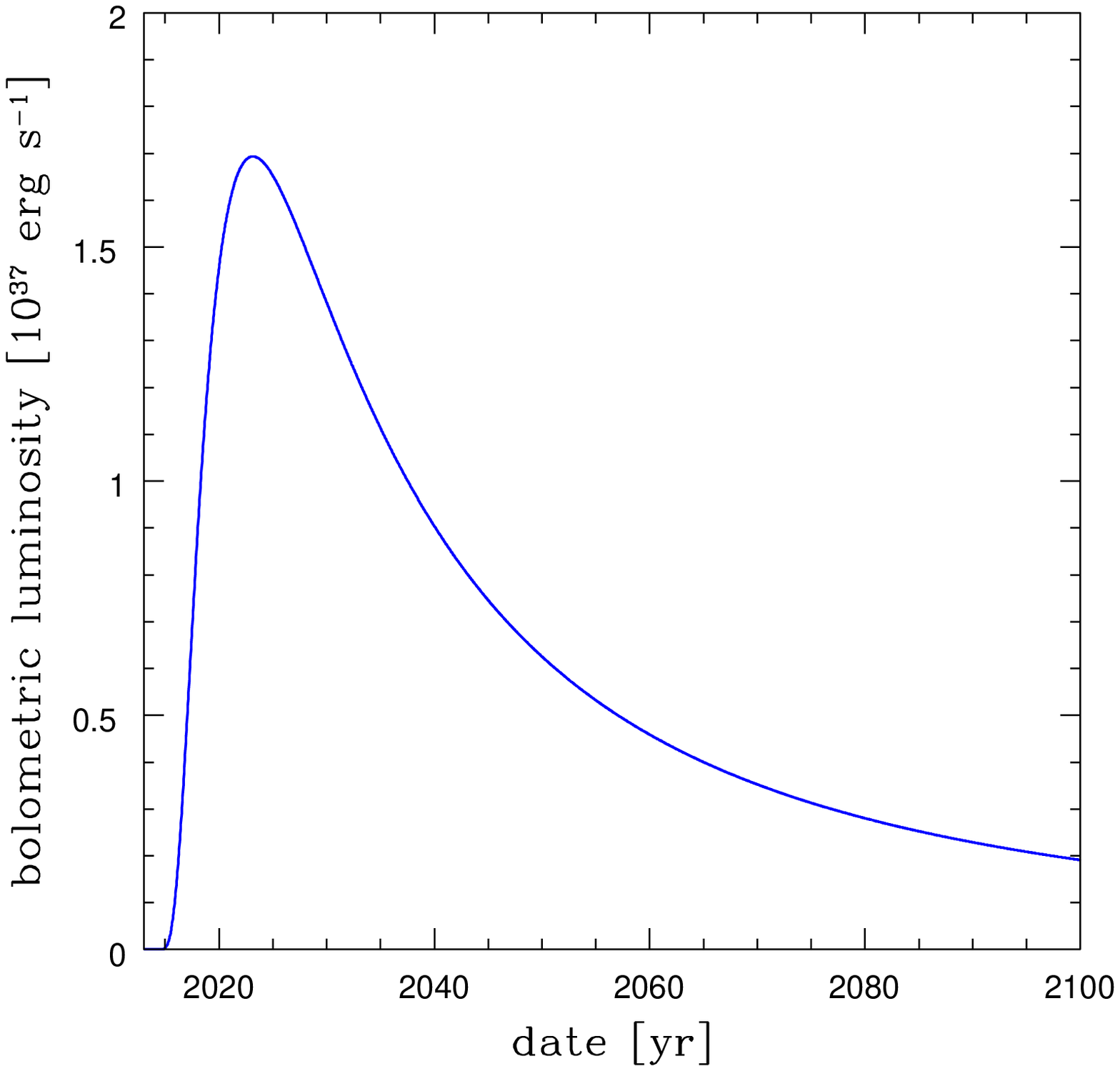}
\hfill
\includegraphics[angle=0,width=0.49\textwidth]{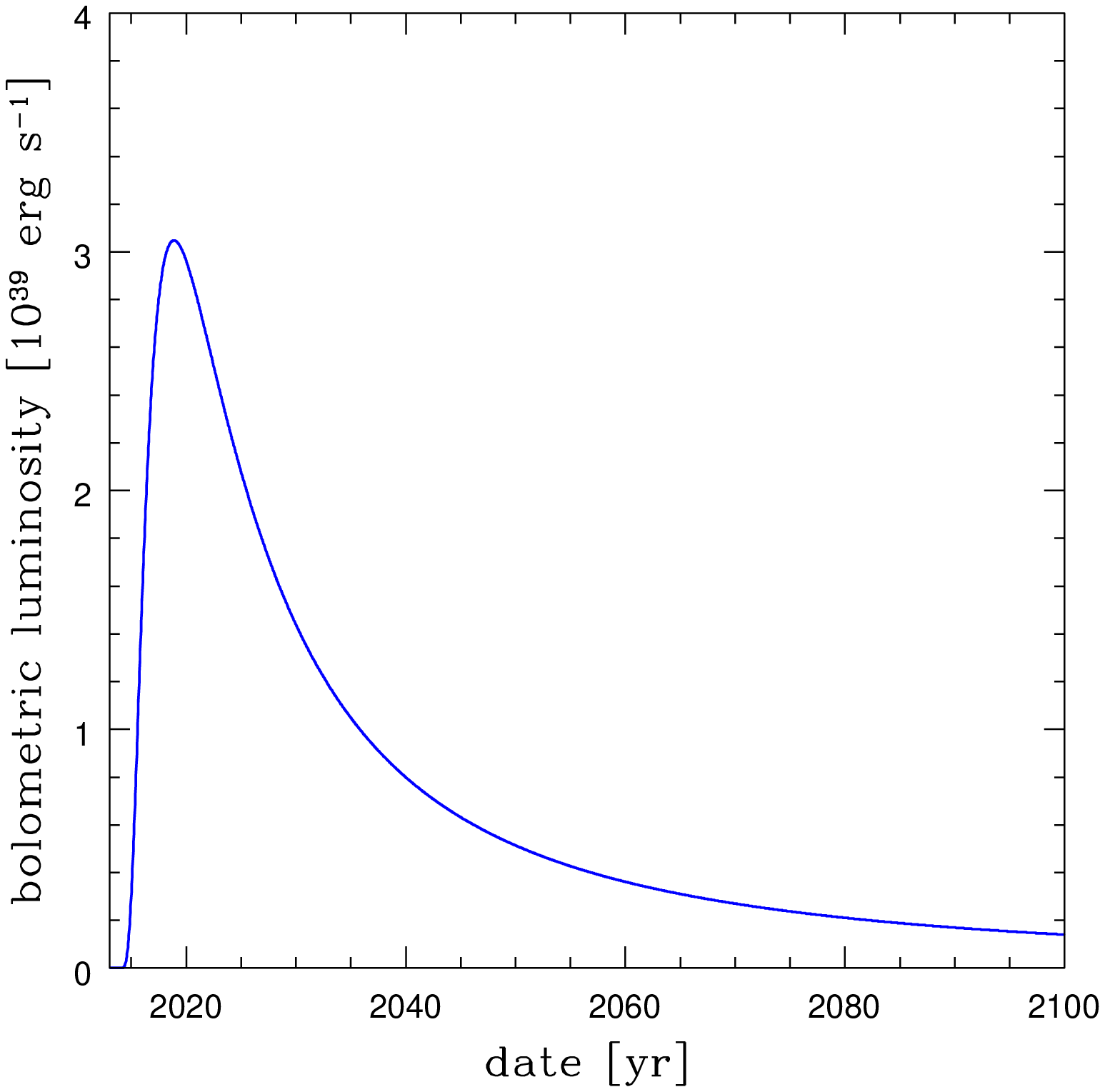}
\end{flushright}
\caption{Exemplary lightcurves generated by a discrete accretion event, as described in the text. Left panel: the expected time dependence of the luminosity excess of Sgr A*  (i.e., the profile of added luminosity) due to a disruption event of a clump of gas with a moderate amount of material, considered as a model for the G2 cloud. Parameters of the model: $t_{\rm visc}=18$~yr, $\eta = 10^{-3}$, $M_{\rm cloud} = 3 M_\oplus$. Right panel: disruption of a massive cloud at smaller pericentre radius: $t_{\rm visc}=10$~yr, $\eta = 10^{-3}$, $M_{\rm cloud} = 3\,10^2 M_\oplus$. Duration of the two events is approximately the same; however, a detailed profile and especially the amplitude of the lightcurve differ significantly from each other.}
\label{fig:G2}
\end{figure*}

It has been reported (Gillessen 2012, 2013) that the G2 cloud is now on an approaching trajectory toward the Sgr A* supermassive black hole. The cloud mass is estimated to be around three Earth masses, $M_{\rm cloud}=3 M_\oplus$, and it moves along an almost perfect parabolic orbit. The predicted pericentre distance of the orbit is about $2200 R_{\rm S}$, although there are several uncertainties in the precise determination of the cloud origin, its properties, and future trajectory (Eckart et al. 2013a,b; Pfiher et al. 2013). We assume that, after an inevitable disruption, the cloud remnants will settle onto a circularization radius around and below pericentre; however, the exact outcome of the forthcoming evolution of the cloud is uncertain and depends critically on its internal structure, which is still under debate. 

If the material gets heated and the thickness of the newly formed torus is $\sim0.3$ of the circularization radius, the expected viscous timescale becomes $\sim t_{\rm visc,1} = 200$~yr. On the other hand, if the torus is somewhat thicker the timescale comes out shorter, about $t_{\rm visc,1} = 18$~yr (with the same circularization radius). The temperature of the material reaches high values, $5 \times 10^7$ K and $5 \times 10^8$~K, correspondingly. The mean level of the transient accretion rate due to G2 material would be roughly $2 \times 10^{-8} M_{\odot}$~yr$^{-1}$ or $3 \times 10^{-7} M_{\odot}$~yr$^{-1}$, respectively, and the mean additional bolometric luminosity becomes $1.3 \times 10^{36}$ erg~s$^{-1}$ or $1.5 \times 10^{37}$ erg~s$^{-1}$. In the first case, the excess of the luminosity will double the current bolometric luminosity, while in the second, the brightening of Sgr~A* will be quite considerable. Any greater increase in luminosity would require greater mass for the accretion event. Our simple model thus gives the result consistent with far more advanced numerical analyses by Schartmann et al. (2012) and Anninos et al. (2012).

In Fig.~\ref{fig:G2} we show the expected evolution of the Sgr A* luminosity caused by the G2 cloud after its disruption and a hypothetical event of mass increase, with the pericentre slightly closer to the black hole ($\sim 1500\,R_{\rm S}$). The radiative efficiency of accretion was set to $\eta = 10^{-4}$ in these examples. We notice that a single event duration is characterized by the viscous timescale, but the lightcurve tail lasts much longer. This is seen in Fig.~\ref{fig:G2}, as well as in the numerical simulations for putative stellar disruption events. The tail is important if the accretion events repeat frequently because it determines the final decay of the observed signal.

\subsection{Case of multiple accretion events}
The probable luminosity of Sgr A* during the past 500 years was predominantly at the level of $10^{39}$ erg s$^{-1}$. It could be due to either an exceptional strong accretion event or a repetitive infall of numerous small clouds, such as an Earth-mass sized G2 cloud. Naturally, what also matters is the efficiency of the conversion of accreted mass into radiation. In the following discussion we set the radiative efficiency at $\eta = 10^{-4}$ and consider different representative possibilities. 

\begin{figure*}
\begin{flushright}
\includegraphics[angle=0,width=0.32\textwidth]{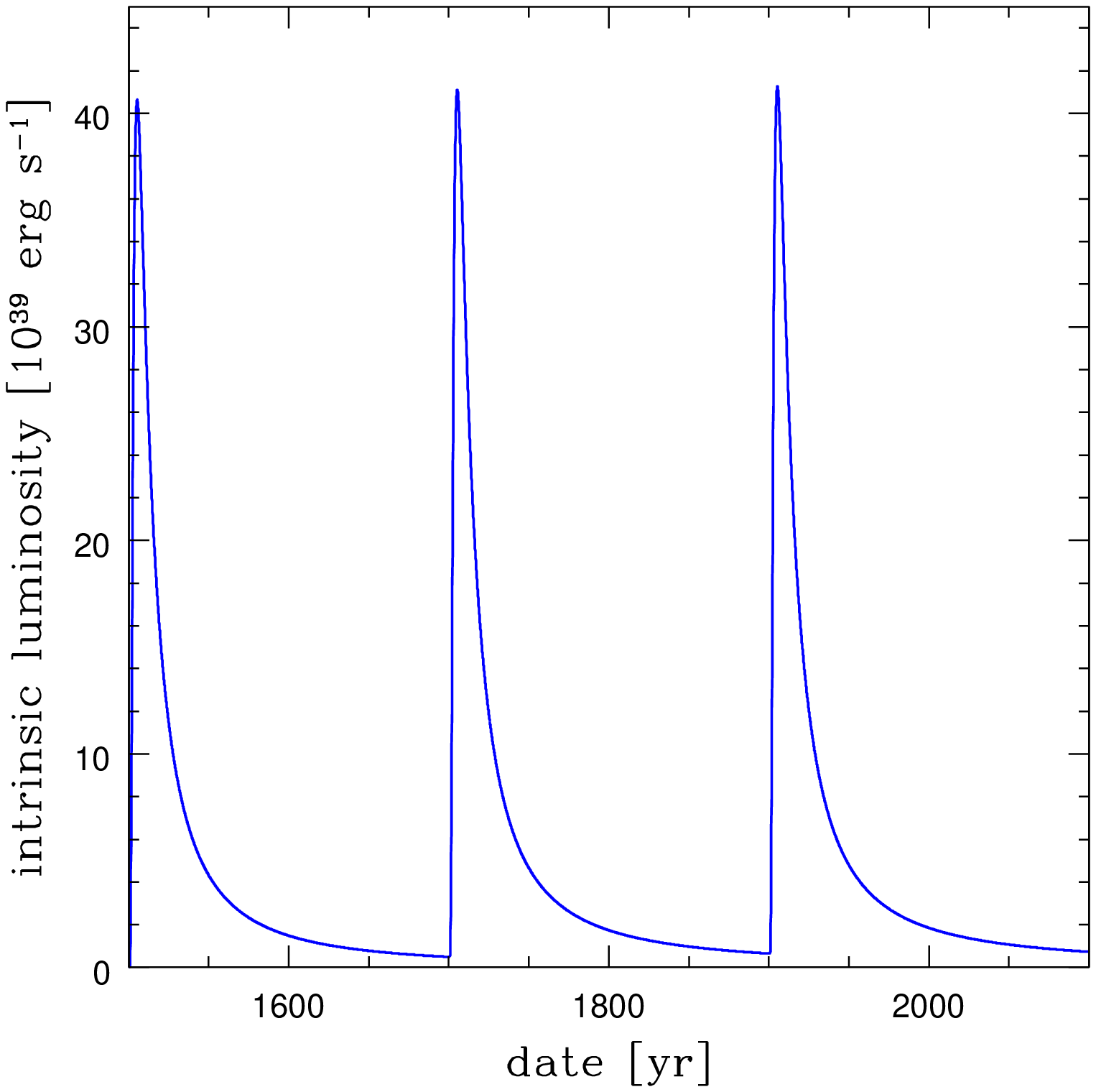}
\hfill
\includegraphics[angle=0,width=0.32\textwidth]{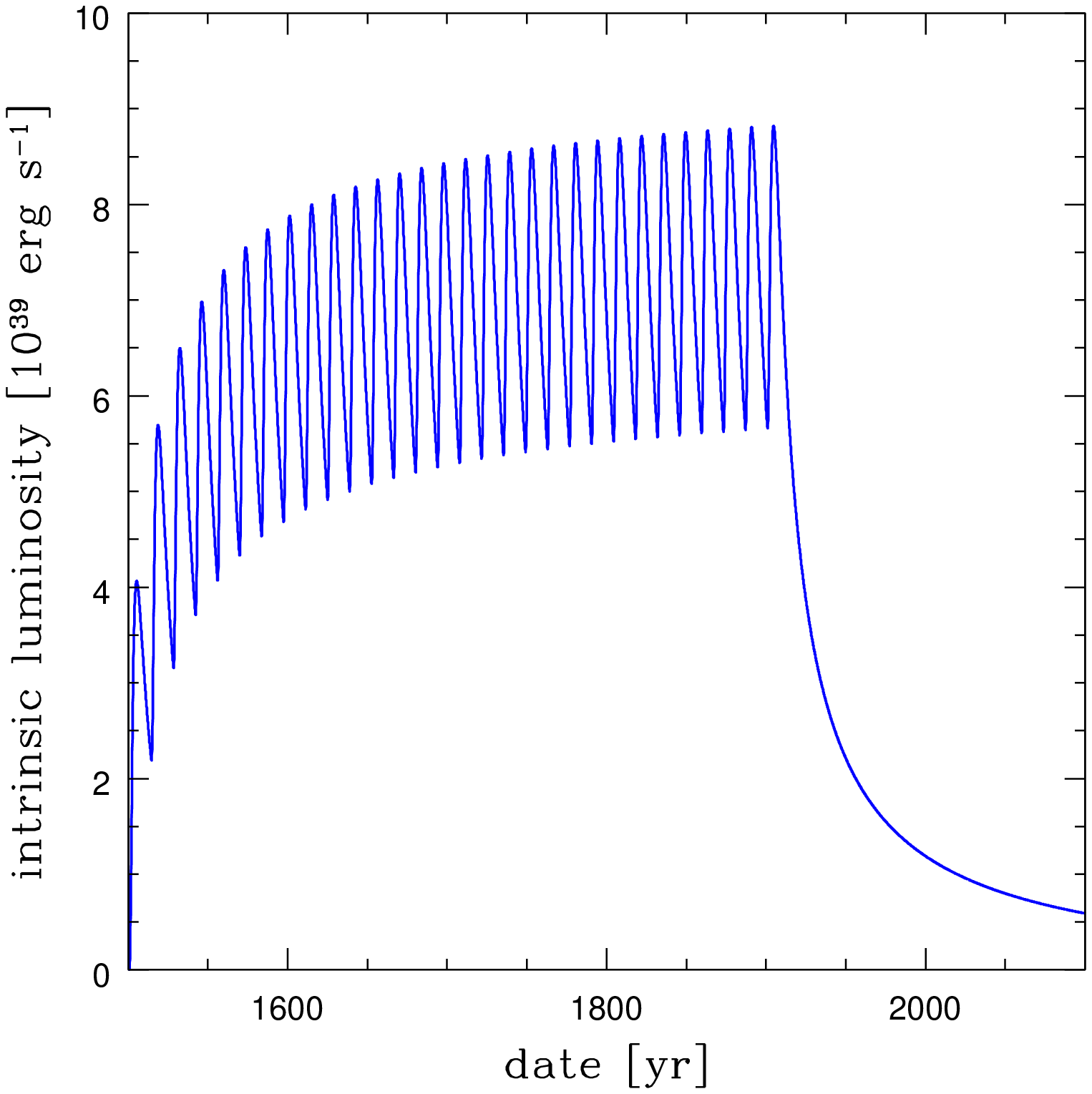}
\hfill
\includegraphics[angle=0,width=0.32\textwidth]{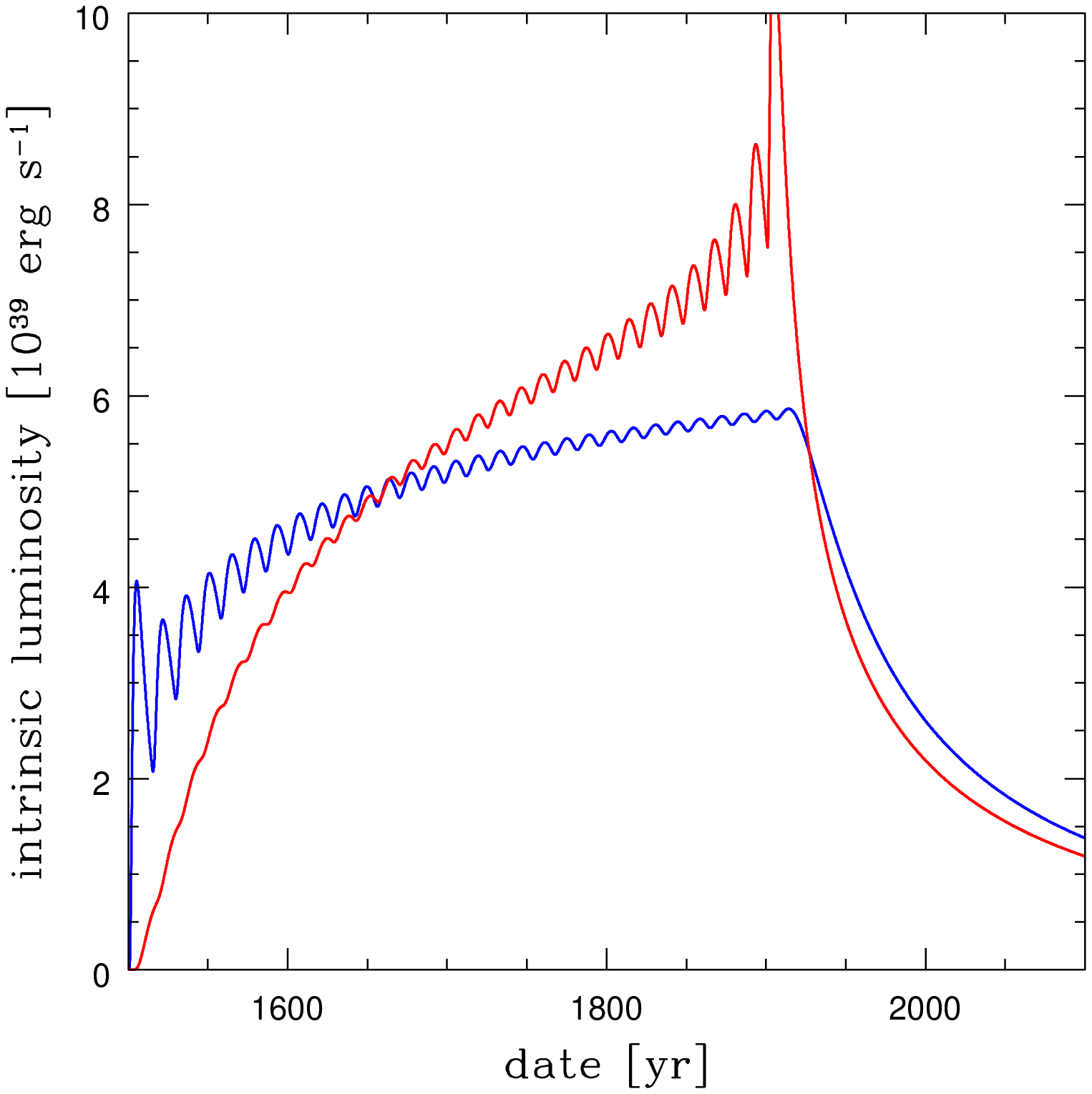}
\end{flushright}
\caption{Examples of qualitatively different possibilities for the luminosity history based on three (left panel) and thirty (middle panel) equal-mass accretion events with the same value of angular momentum. Parameters $t_{\rm visc}=10$~yr, $\eta =10^{-3}$, $M_{\rm total} = 0.03 M_{\odot}$ in both cases. In the right panel we consider thirty equal-mass events, but now with the angular momentum changing gradually over the whole duration. The change is assumed to be linear, namely, the rising profile (blue curve with relatively small variations) vs.\ decreasing profile (red curve, with a sharp rise followed by the rapid decline) between the corresponding values of $t_{\rm visc}=10$~yr and $t_{\rm visc}=50$~yr, respectively.}
\label{fig:multi}
\end{figure*}

Provided that all the accreting clumps reach a similar circularization radius, the overall shape of the lightcurve depends mainly on the temporal distribution of the events. We consider equally spaced events with the fixed total accreting mass roughly corresponding to the mentioned luminosity during the past few hundred years. The simplifying assumptions allow us to show the expected effects more clearly, but they can be easily relaxed for an astrophysically more realistic discussion. 

In Figure~\ref{fig:multi} we show examples with three and thirty events. Time moments of these events are set in such a way that they cover the period from the year 1500 to 1900. In the first example, a single flare profile is well preserved. In the second case, regular luminosity fluctuations persist but they are less than a factor 2 intense, superimposed on the overall luminosity rise due to event overlapping. 

The angular momentum of the clumps may also span a broad range of values. Again, we consider thirty accretion events, spaced regularly as in the previous example, and we perform two exercises: in the first case the angular momentum of the subsequent clumps rises, while in the second it decreases. In both we assume the linear distribution in angular momentum, and the maximum and minimum values correspond to the maximum and minimum values of the viscous timescale of 10 and 50 years. In both cases the variability amplitude decreases in the main part of the lightcurve due to a strong overlap of events. We just point out that these plots illustrate the role of the frequency of repetitive accretion events and of the angular momentum distribution of the accreting clouds on the overall shape of the lightcurve. Particularly constraining is the rate of the final drop of the luminosity when accretion stops.

\subsection{Mechanism generating the intrinsic lightcurve}
\label{sect:SgrA_curve}
Several authors have attempted to reconstruct the history of the Sgr A* activity. We discuss this interesting problem again in terms of the source lightcurve during the recent era and set constraints that appear to be consistent with our model. To this end we have adopted data from the most recent study by (Ryu et al. 2013). The data is particularly interesting for modelling since the results are based on three-dimensional reconstruction of the positions of the molecular clouds. The lightcurve exhibits a typical variability during the last few hundred years by about a factor three. The rapid decline by about four orders of magnitude happens in the course of the last 60 years. However, we also show the data set from Capelli et al. (2012). These data points represent the 2--10 keV luminosity of the source. Ryu et al. (2013) give the data directly in this format. Points in the Capelli et al. (2012) are given in 1--100 keV flux, and these were converted to 2--10 keV flux using again the mentioned models by Moscibrodzka et al. (2012). The correction moves the points not more than by a factor 2.2.

We first consider the case of constant accretion efficiency $\eta \simeq 10^{-3}$, independent of the Eddington luminosity ratio, $L_{\rm bol}/L_{\rm Edd}$ and constant conversion factor between the bolometric and the X-ray luminosity. In this case the rapid decline imposes a stringent limit on the viscous timescale. The most rapid X-ray decay decrease in the Ryu et al. (2013) data is by a factor of $10^4$ in only 60 years. Considering the asymptotic behaviour of the flow with the luminosity decreasing as $\tau^{-(1 +\mu)}$, we obtain, in agreement with eq.~(\ref{eq:mdot1}), the viscous time $t_{\rm visc} = 60/10^{4/(1 + \mu)}$~yr. This leads to $t_{\rm visc} = 0.24$~yr for $\mu = 2/3$. In the Capelli et al. (2012) data, the most rapid decrease is by a factor of 100 in only 14 years, suggesting the shortest timescale of $t_{\rm visc} = 0.88$~yr.

\begin{figure*}
\begin{flushright}
\includegraphics[angle=0,width=0.45\textwidth]{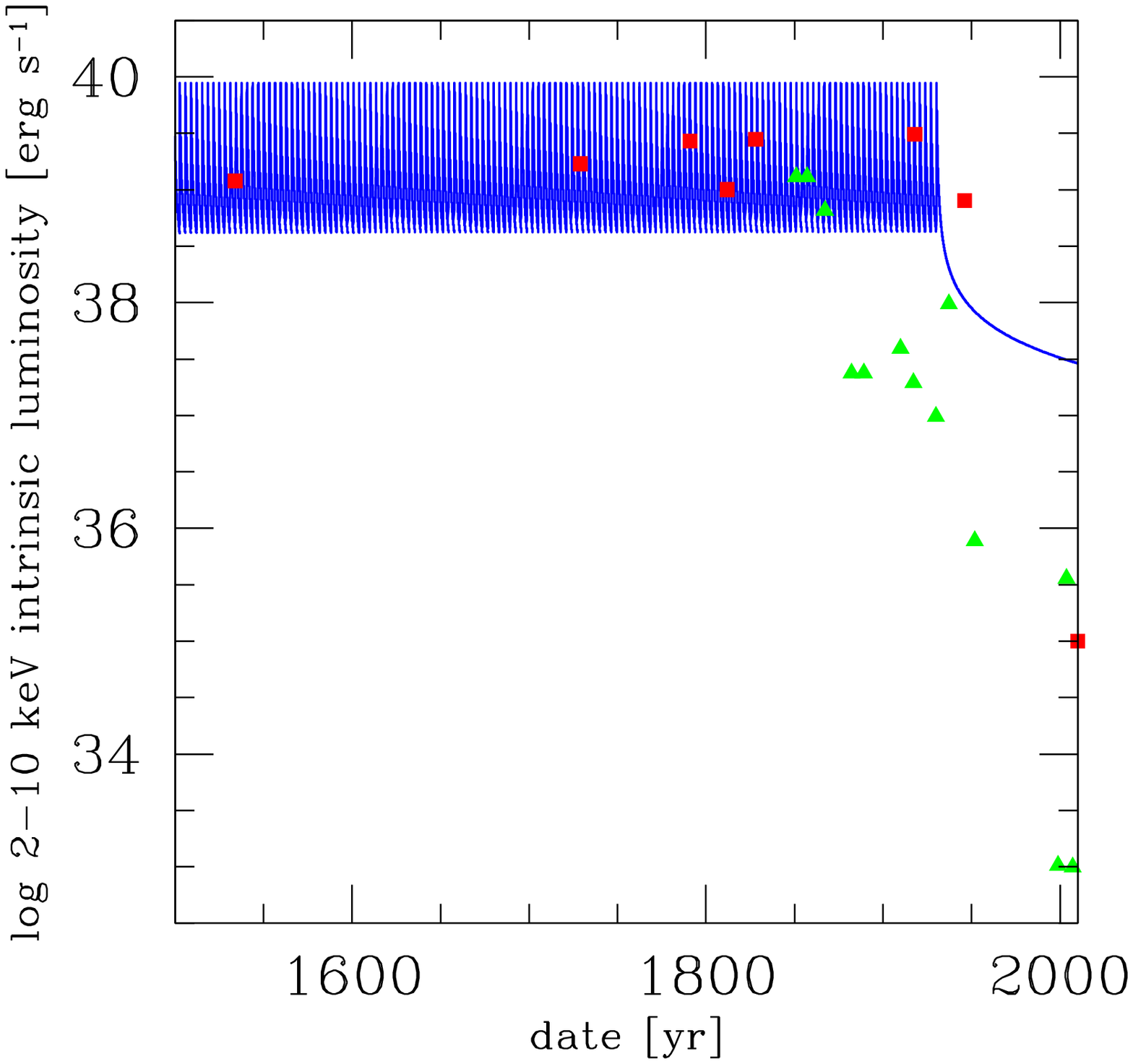}
\hfill
\includegraphics[angle=0,width=0.45\textwidth]{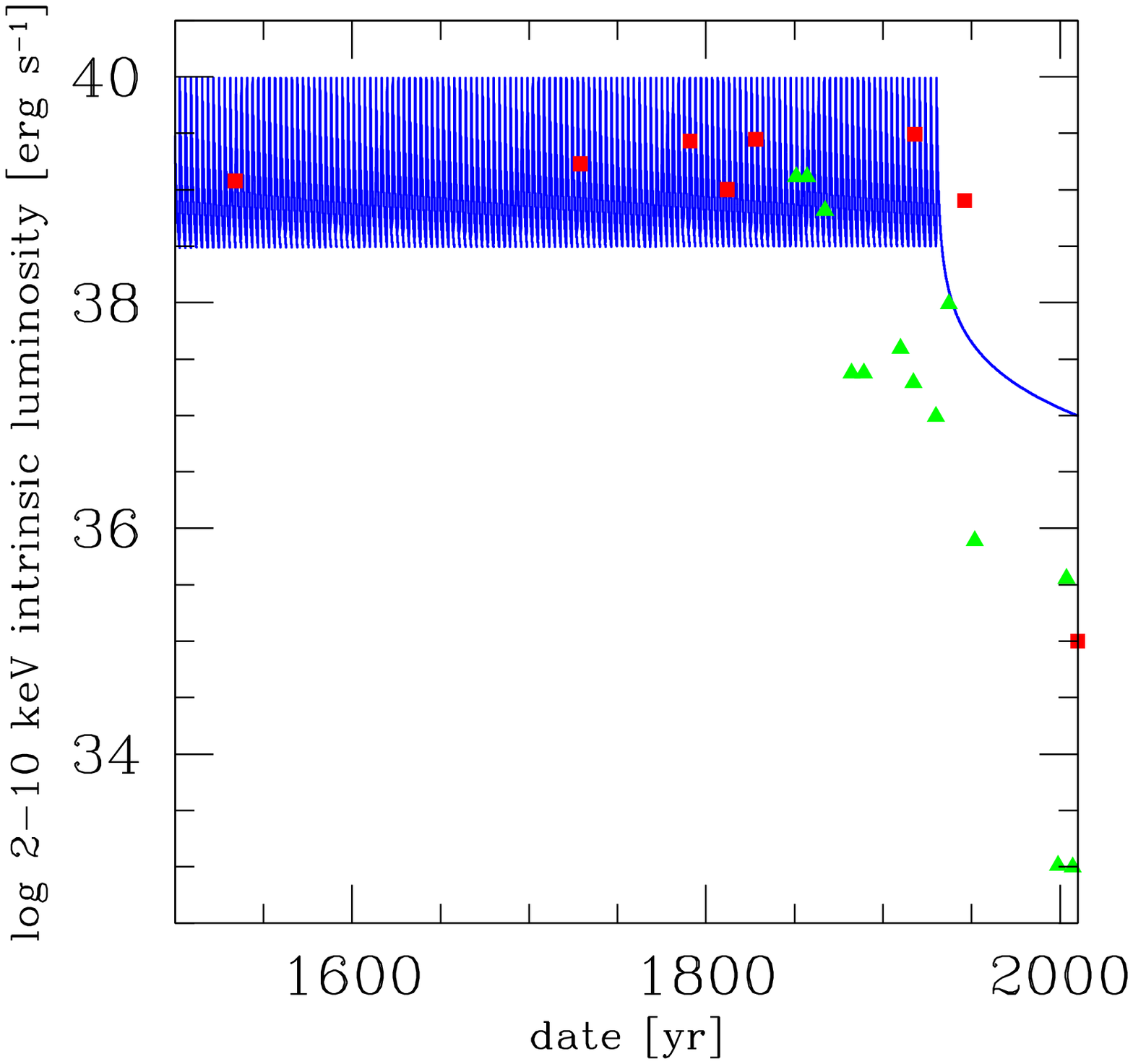}
\hfill
\includegraphics[angle=0,width=0.45\textwidth]{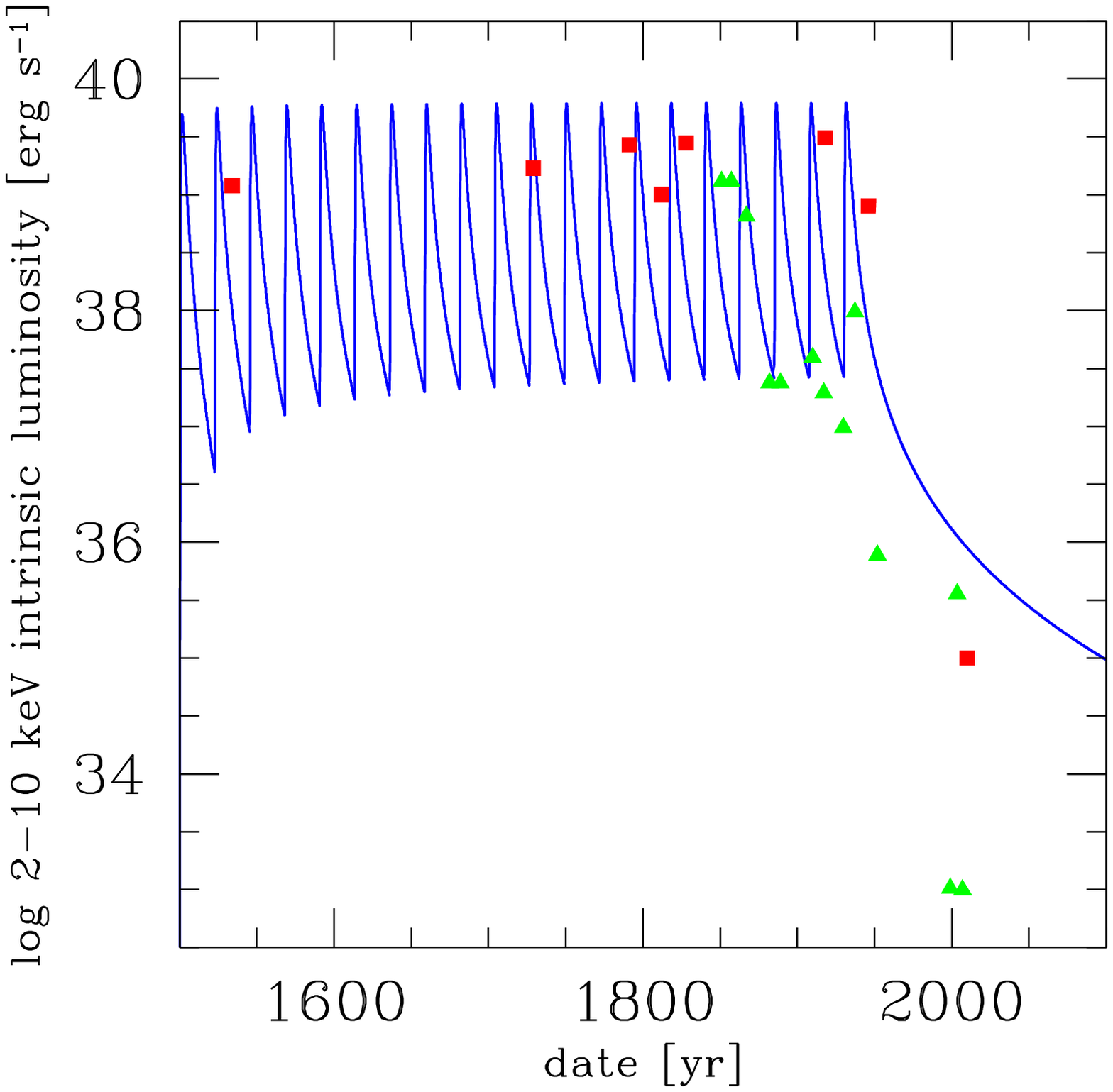}
\hfill
\includegraphics[angle=0,width=0.45\textwidth]{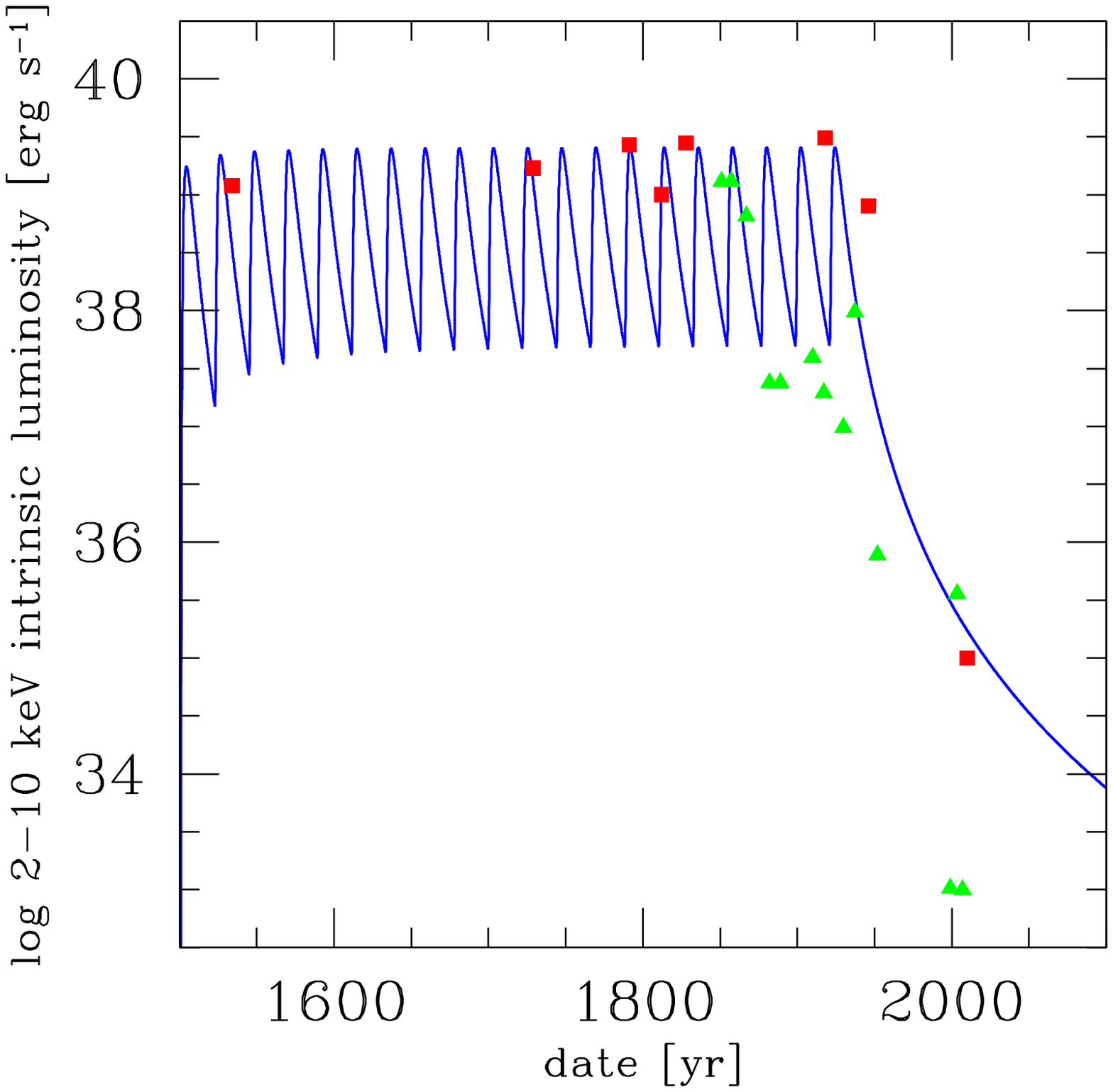}
\end{flushright}
\caption{Four cases of the predicted luminosity history for different values of the model parameters representing the decay and the variability level with a reasonable mean luminosity and the range in its variation. Constraining data values are shown by red squares (based on X-ray data from Ryu et al. (2013), plotted here without the errorbars) and green triangles (based on Capelli et al. 2012). The model lightcurves are plotted by a blue continuous curve. The interplay of model parameters determines the level of mean luminosity, frequency of the oscillations, and the rate of final decline at the moment when the accretion ceases. Top--left panel: $M_{\rm tot} = 0.1 M_{\odot}$, $t_{\rm visc}= 0.24$~yr, $N = 200$, $\mu = 2/3$, $\eta = 10^{-3}$, $\eta_X= 0.1$. Top--right panel:  $M_{\rm tot} = 0.1 M_{\odot}$, $t_{\rm visc}= 0.24$ yr, $N = 200$, $\mu = 2/3$, $\eta = 10^{-3}$, $\eta_X$ from Eq.~\ref{eq:bolo_cor}. Bottom--left panel:  $M_{\rm tot} = 0.025 M_{\odot}$, $t_{\rm visc}= 3$ yr, $N = 20$, $\mu = 2/3$. Bottom--right panel: $M_{\rm tot} = 0.075 M_{\odot}$, $t_{\rm visc}= 3$~yr, $N = 20$, $\mu = 1.3$. In the last two cases, $\eta$ was determined from eq.~(\ref{eq:eta}) and $\eta_X$ from eq.~(\ref{eq:bolo_cor}).}
\label{fig:multi_500}
\end{figure*}

The total accreted mass is a fraction of the solar mass ($M_{\rm tot}\simeq0.1 M_{\odot}$ in our exemplary solution). The tail from multiple episodes slows down the decay of the emission further. We would be able to roughly reproduce the lightcurve fall off and the general level of variability only for a viscous timescale as short as $t_{\rm visc} = 10^{-3}$ yr (i.e. below 1 day), so this short timescale is not a realistic value. When we use the timescale of 0.24 yr, suggested by the simple analysis above, the decay is far too slow in comparison with the data due to an extended period of activity (see Fig.~\ref{fig:multi_500}, upper left). The situation improves if we use the bolometric to X-ray conversion flux according to Eq.~(\ref{eq:bolo_cor}) because the decrease in the bolometric luminosity is additionally accentuated by the decline in the conversion factor, but not enough to match the observed trend (see Fig.~\ref{fig:multi_500}, upper right).

We can consider higher values of the coefficient $\mu$. A steeper decay in the accretion rate (even as steep as $t^{-2.3}$) has been seen in numerical simulations of the stellar disruptions  (e.g. Guillochon \& Ramirez-Ruiz 2013, see their Fig.~5). The analytical requirements for the fast time decay is somewhat less stringentfor $\mu = 1$, $t_{\rm visc} \simeq 0.6$ yr. We can then reproduce the observed trend in Sgr~A* numerically for the following parameters, the total mass $0.2 M_{\odot}$, $t_{\rm visc} = 0.005$ yr, the number of events $10^4$, so the mass of a single event is about ten Earth masses. Such a model still does not look promising because the viscous timescale is again far too short, the circularization radius is only $\sim 9 R_{\rm S}$, and the required number of events is very large. 

To resolve the above-mentioned problem with a rapidly decaying lightcurve, we consider the radiative efficiency to be a function of the Eddington rate given by the Eq.~(\ref{eq:eta}). In this case the decrease in the accretion rate translates into an even faster drop of observed luminosity. This opens up the possibility of reproducing the observed decay in activity with less extreme values of the model parameters. If the standard $\mu = 2/3$ value is used, the model with the decay timescale of 3~yr does not yet match the observed luminosity decrease (e.g., Fig.~\ref{fig:multi_500}, bottom--left panel); however, if we couple the variable efficiency case with a steeper decay (as suggested by the modelling with $\mu = 1.3$), the solution with the viscous timescale of three years represents the observed trends reasonably well (see Fig. \ref{fig:multi_500}, bottom-right panel). Also, the number of accretion events is now considerably lower, $\sim20$. 

\begin{figure*}
\begin{flushright}
\includegraphics[angle=0,width=0.49\textwidth]{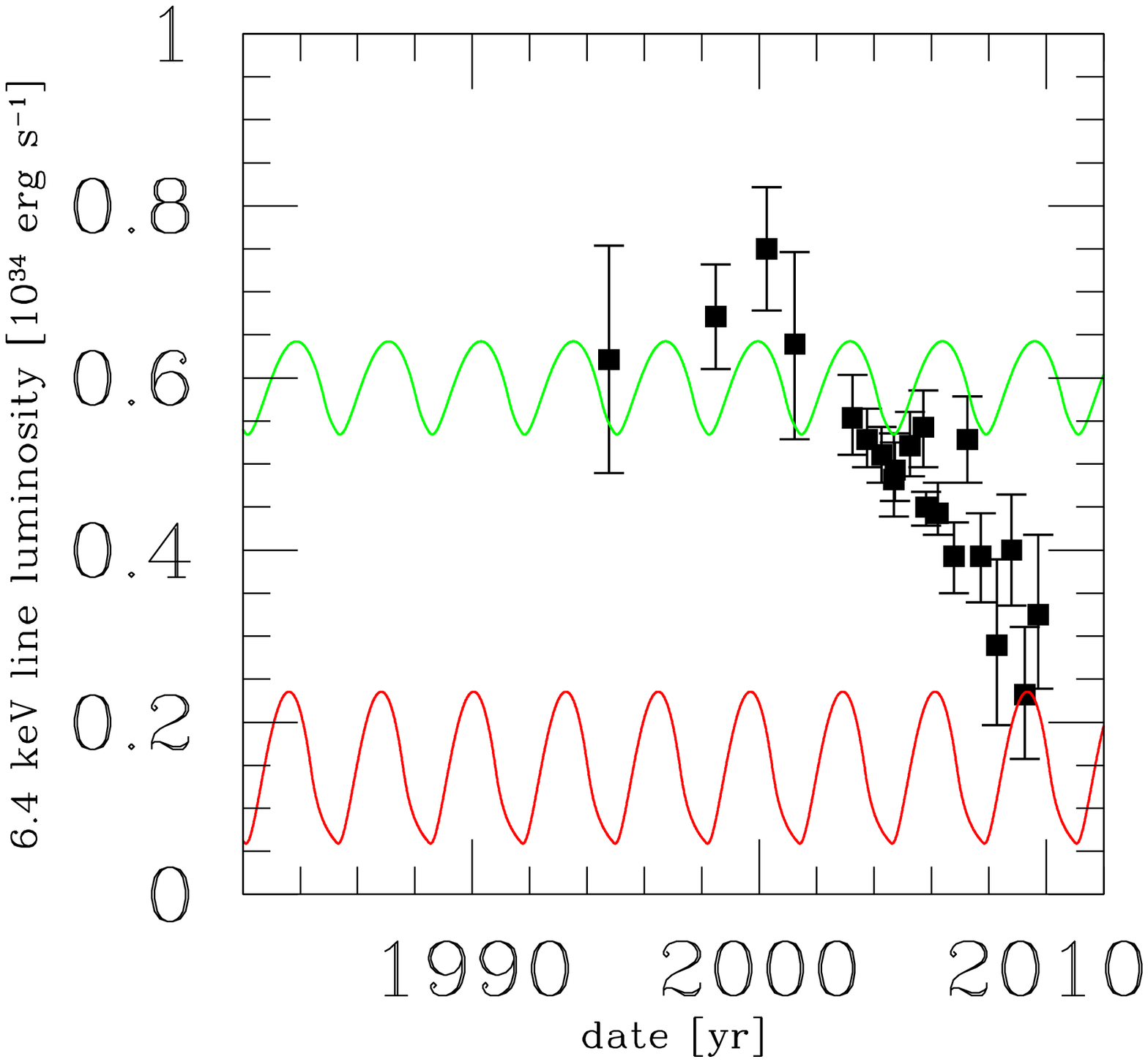}
\hfill
\includegraphics[angle=0,width=0.49\textwidth]{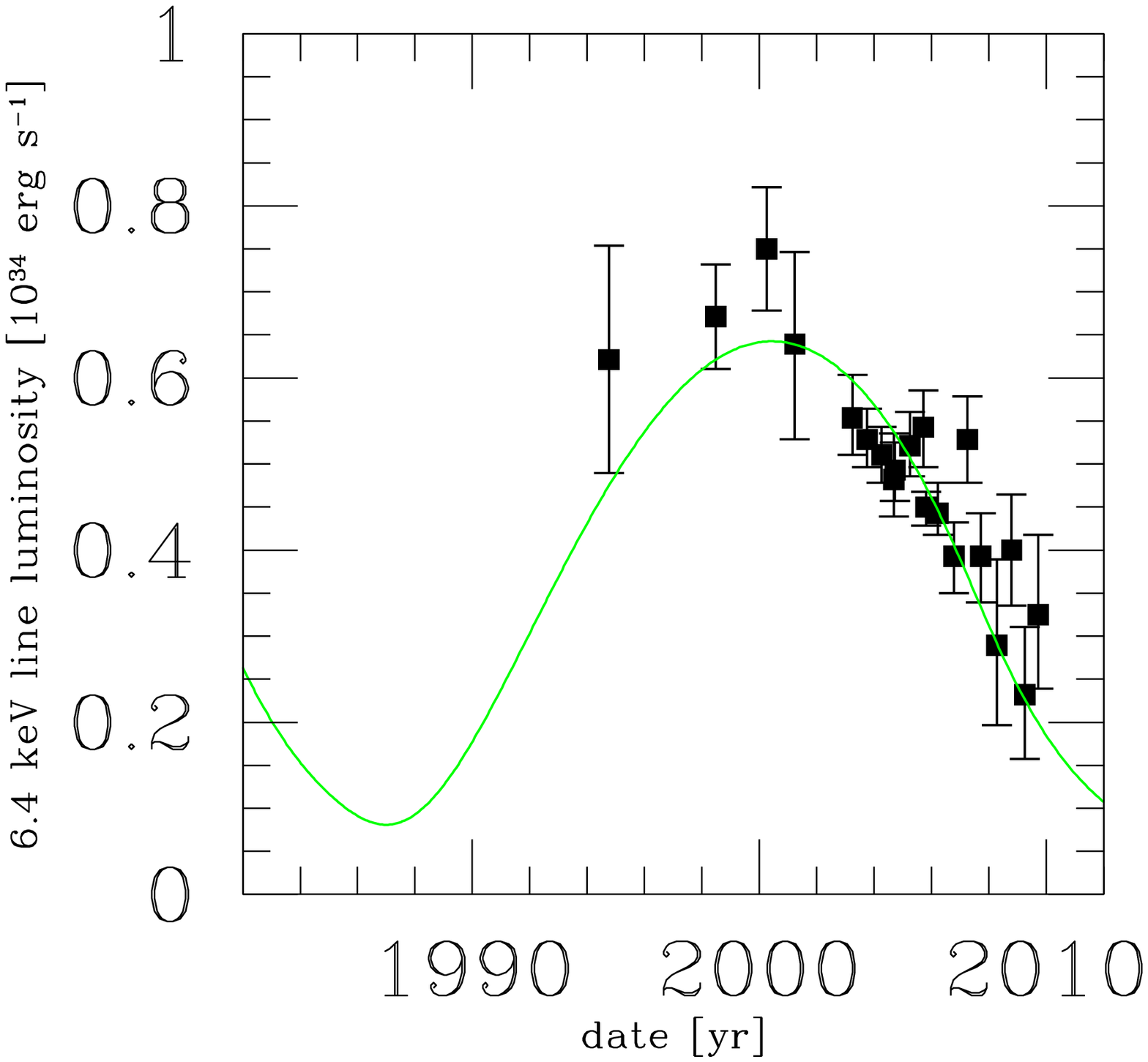}
\end{flushright}
\caption{Test examples of the time dependence of the iron K$\alpha$ reflection spectral line flux from the model of Sgr B2 cloud reprocessing (continuous lines), together with the data points from Yu et al. (2011; shown by squares and the corresponding measurement error bars). The constant density model was used with parameters as follows. Left panel: the accretion model as in the top--left panel of Fig.~\ref{fig:multi_500}, $R_{\rm cloud} = 0.25$~pc (red line) and $R_{\rm cloud} = 0.50$~pc (green line); obviously this example does not produce a reasonable solution because either frequency or the amplitude of oscillations do not correspond with the data. Right panel: as in the bottom--right panel of Fig.~\ref{fig:multi_500}, $R_{\rm cloud} = 2.3$~pc; the latter example provides significantly better agreement with the data for which the formal $\chi^2$ test converges to an acceptable fit.}
\label{fig:B2}
\end{figure*}

Indeed, the model cannot easily represent the Sgr ~A* luminosity history with sufficient accuracy. Especially the rapid decay phase, by many orders of magnitude, implies short viscous timescales, or equivalently, low values of the circularization radii. Today the X-ray emission is at the level of $10^{33}$--$10^{35}$ erg~s$^{-1}$, in the quiescent or flaring state (e.g. Baganoff et al. 2001; Porquet et al. 2003; Genzel et al. 2010, and further references cited therein). The fit shown in the right-hand panel of Fig.~\ref{fig:multi_500} is close to the required profile, but it is neither unique nor perfect. Additional constraints can be obtained from studying the reflection by B2 and C1 clouds.

\subsection{Reprocessing by molecular clouds B2 and C1}
The Sagittarius B2 (\object{Sgr B2}) complex is a prominent example of a large molecular cloud system situated in the eastern part of an active star-forming region at $\simeq100$~pc from Sgr~A*. Chandra images of Sgr B2 cloud (Murakami et al.\ 2001; Ryu et al. 2009) and the combined evidence of the variability of the region from different X-ray instruments (Suzaku, XMM-Newton, Chandra, and ASCA; see Inui et al.\ 2009; Nobukawa et al. 2011, and references cited therein) demonstrate the complex morphology of the reflecting environment with enhancements of density in multiple cores. 

The fluorescent iron K$\alpha$ emission of Sgr B2 cloud should follow the general trend of Sgr A* intrinsic luminosity with a delay and smearing of the signal due to the finite distance and size of the cloud. Currently this emission exhibits a fading trend, but we can only see a small fraction of the lightcurve. This trend may reverse in the next several years, as implied by higher luminosity of the clouds that are located at closer distance from Sgr A*. Therefore we cannot model the trend of B2 luminosity arbitrarily; instead, we have to use the long timescale lightcurve (cf.\ sect.~\ref{sect:SgrA_curve}), select a specific period, and adjust the beginning of the burst sequence in such way that the flare appears at the right position. 

We concentrated on the best studied case of Sgr B2 cloud and adopted the compilation of data points by Yu et al. (2011). We took the specific examples of accretion pattern presented in different panels of Fig.~\ref{fig:multi_500}, calculated the lightcurve for the reprocessed emission, and then compared the B2 reflection data to the suitable fragment of the lightcurve. Since the delay of the signal from the B2 cloud is about 280 years (Ryu et al. 2013), the currently observed reflection (covering the period $\sim 1995$--$2010$) develops from fragments of the intrinsic lightcurve produced as early as in 1715--1730.  We need to only slightly adjust the beginning of the sequence of accretion events to reproduce the peak at the right place. 

We set up several test examples in order to understand the trends present in the model. In the case of infall parameters shown in the top panels of Fig.~\ref{fig:multi_500} there is virtually no possibility to model the reflection. The number of accretion events comes out as far too high; namely, several events in just two decades owing to the very short viscous time in these solutions. In the case of a small size of the reprocessing cloud, we simply see a wave in the lightcurve, instead of a gradual slow decay trend, and the amplitude of the 6.4 keV flux is too low because small fraction of X-ray photons are intercepted and reprocessed by the cloud (see Fig.~\ref{fig:B2}, left panel). If the cloud size is set larger, the mean luminosity is comparable to the observed one, but the the lightcurve is considerably smeared, with the amplitude much smaller than the currently observed flux changes. In principle, it is possible to imagine that there is a temporary trend toward decreasing or increasing the mass of the event with time, but the timescale itself and the corresponding circularization radius do not produce realistic values. On the other hand, the model in the bottom right-hand panel of  Fig.~\ref{fig:multi_500} (which adopts the realistic description of the accretion efficiency and includes the bolometric to X-ray conversion and the increased value of parameter $\mu$) represents the behaviour of the B2 reflection very well. In fact, we only had to adjust the starting point of the sequence and the cloud size. 

We note that here we only show an example under the assumption of constant density; in this case, we checked that the case of a more extended cloud with an rarefied envelope does not improve the solution significantly. The formal best fit was obtained with the diameter of 2.8~pc, but then the first four points fall far above the curve. Therefore we consider the cloud of the size 2.3~pc as a better representation of the data (see Fig.~\ref{fig:B2}). 

\begin{figure}
\includegraphics[angle=0,width=0.49\textwidth]{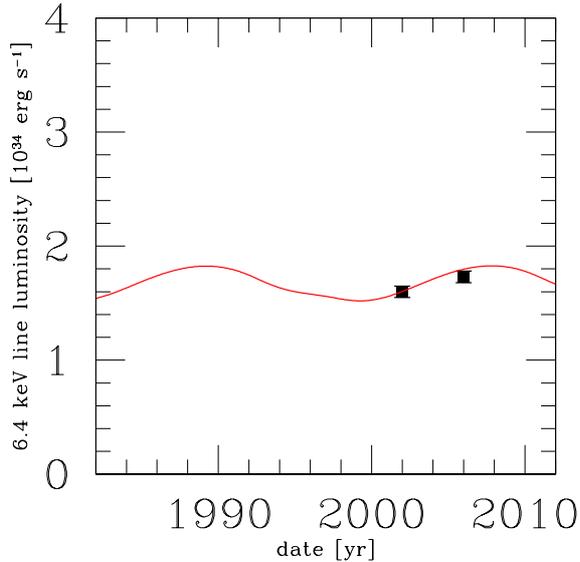}
\caption{The modelled time dependence of the K$\alpha$ luminosity of C1 cloud (green line) and the data points from Ryu et al. (2013; filled squares) with measurement 
error. Model parameters: accretion model as in Fig.~\ref{fig:multi_500} (right panel) and Fig~\ref{fig:B2} (right panel),  $R_{\rm cloud} = 2.6$~pc. Despite so few data points, it is important that the overall normalization of the flux comes out consistent with the model and does not require further tuning of the model parameters.}
\label{fig:C1}
\end{figure}

We do not expect that accretion events are strictly periodic, so we do not suggest that the past data predict a future rise of the B2 luminosity. The separation between events is more likely random, and the next event may be delayed by several years. On the other hand, the luminosity trend of the clouds at different radii close to the black hole indicates quite convincingly that this will indeed happen. With measurements of more clouds than those given in the recent paper (Ryu et al. 2013), we should be able to recover the time moments of separate events. However, it is interesting to note that the simple periodic accretion with constant angular momentum and constant amount of material involved in these events already reproduces the overall profile of the reflection seen from  Sgr B2 very accurately. 

Next, we also included the variations in the iron line flux measured from \object{Sgr C1} cloud (Ryu et al. 2013). The delay and the true distance of this cloud from Sgr~A* are well constrained, so we can employ the same model as explained above for the modelling Sgr B2. We varied the cloud size, and by increasing it considerably we achieved both the flux enhancement and a shallower variation profile, as required by the data. However, there are only two data points available, and so the fit is ambiguous (Fig. \ref{fig:C1}). Nevertheless, it is important that both the luminosity and the shallow profile of the variation can be reproduced with only slight adjustments: the event sequence was shifted by 3.5 yr (i.e. within an error of the timedelay measurement for B2 and C1), but otherwise the flux was calculated for the system geometry assuming the same density everywhere.

As a second example we calculated the core and envelope set. Since the number of data points is not large, we had to constrain the number of free parameters. We thus fixed the core radius at 0.25 pc, as adopted by Odaka et al. (2011). In this case the best fit was achieved for an envelope radius of 2.6 pc, so larger than in the single constant density cloud (2.3 pc), but otherwise the fit quality did not improve considerably.

\begin{figure}
\includegraphics[angle=0,width=0.49\textwidth]{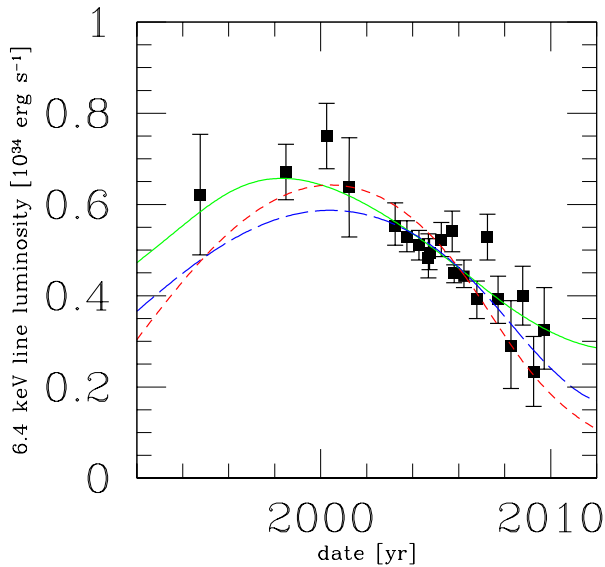}
\caption{The preferred description of the time dependence of Fe K$\alpha$ declining luminosity of the B2 cloud. (i)~Short-dashed (red) line: the best fit for constant density model of the cloud; (ii)~long-dashed (blue) line: two-component constant density model of the core and envelope; (iii)~thick continuous (green) line: dense core with a radially decreasing density of the envelope. This final model gives the most precise representation of the data and also seems to describe the most realistic situation.}
\label{fig:summary}
\end{figure}

Finally, we used the envelope with the density decreasing as $R^{-1}$, fixing the core again at 0.25 pc. This gave the best representation of the data. The outer radius of the envelope came out equal to 3.5 pc, larger than in the constant density case. This solution  is also more consistent with the observational estimates of the B2 size. We show the comparison of the three cases in Fig.~\ref{fig:summary}. 

\section{Discussion}
\label{sec:discussion}
Sgr A* had exhibited a significantly enhanced level of activity during the past few hundred years, and it experienced a significant decrease in the activity over the past several decades. In this paper we considered a simple model that aimed to describe the nature of these changes and reveal the peculiarity of the long-lasting period. The model points toward a possible origin for the enhanced activity in terms of accretion of individual gaseous clouds. 

We parameterize the activity in the form of the sequence of events with a definite moment of start date and the end date. All the accreting clumps settle onto a circularization radius, spread during the viscous time, and subsequently feed the black hole for a certain period of time. We followed the viscous evolution of the events and calculated the intrinsic luminosity of the central black hole from the (time-dependent) accretion rate across the inner boundary of the disc. We showed the most simplistic case of events generated with equal spacing in time; however, a modification to a stochastic generation of clouds is straightforward and it does not create any qualitative difference. Also, the amount of accreted material and the corresponding angular momentum of each clump are free parameters that can be set arbitrarily within a certain range of allowed values. However, if we assume events of equal mass, the shallow Sgr A* lightcurve can only be represented by assuming that the angular momentum in all events stays constant. 

Quantitatively, the model should reproduce two main properties of the lightcurve: (a) the current decay in the activity after the last event, and (b) the shallow variability during the bright state. The latter has been implied by the slow variability of light reprocessed by the surrounding molecular clouds. The rapid decline in the lightcurve indicates a short viscous timescale. However, this leads to rapid and significant variability in the past (changing the intrinsic flux by orders of magnitude), unless the number of events was very large. If this was the case, all the variability would have been strongly smeared during the reflection process, and the molecular cloud images should have exhibited a constant flux. A very short decay timescale is also a problem from the point of view of the angular momentum requirements: the material must have settled directly onto a circular orbit very close to the black hole at only a few Schwarzschild radii, which does not seem to be realistic.

On the other hand, in the case of  the rising trend of the angular momentum, the lightcurve would have been flat but its normalization too high to reproduce the sudden decrease in the activity seen in the data. If the angular momentum of the later events is lower the overall luminosity rises gradually, and this also does not conform to the expectation. A flat profile seems to be more likely (cp.\ Fig.\ \ref{fig:multi}). Of course, assuming that the mass of the accretion events is coupled in some peculiar way to the angular momentum, we can always achieve the flat profile of the lightcurve; however, the Ockham razor argument can be used against such an approach. To summarize, given the quality of available data, we restricted ourselves to those events spaced equally in time, mass, and angular momentum parameters as an adequate approach to the problem.

Our modelling of the viscous evolution of the clumps arriving down to the black hole suggests that two additional assumptions are needed: (i) the radiative efficiency of the flow depends on the Eddington ratio, and (ii) the viscous decay should proceed somewhat faster than the canonical $\propto t^{-5/3}$ profile, which is frequently considered in the context of material  spreading viscously after a tidal disruption event. The first aspect can be justified: the observational data for accreting black holes in binaries suggest such a trend (Narayan \& McClintock 2008). This coupling makes the requirements for the drop in the accretion rate less stringent because it implies an additional decrease in the radiative efficiency $\eta\equiv\eta({\dot{M}}(t))$. The latter modification is also supported by recent numerical simulations (Guillochon \& Ramirez-Ruiz 2012). It appears that the values of the power-law index characterizing the decay timescale may spread somewhat around the canonical value in both directions, i.e., towards a faster, as well as slower, decline of the lightcurve (for a different example, see e.g. Krolik \& Piran 2011). This reflects various details of each particular case, such as the exact location of the disruption event with respect to the tidal radius that determines the rate of draining of the accretion disc. 

With the two above-given assumptions we reproduce the overall activity of the source by a sequence of infalling clumps accreting over a long period spanning from about $1400$ till $1930$s. The total mass is estimated to be $M_{\rm tot}\simeq0.15 M_{\odot}$, and the viscous timescale about $3$ years (the latter value corresponds to the circularization radius $R_{\rm circ}\simeq 700 R_{\rm S}$). 

The same approach to the intrinsic lightcurve was used to fit the data for two representative clouds, B2 and C1. For these we took the observed K$\alpha$ lightcurve, together with 3D positions needed to determine the distance between each of the clouds and Sgr A*, as well as the timedelay between the intrinsic and the reprocessed emission.  We note that complete information on the spatial location of the clouds is needed to normalize of the reprocessed flux. We employed values for B2 (Terrier et al. 2010; Yu et al. 2011) and two additional points for C1 (Ryu et al. 2013).

Modelling the B2 X-ray reflection required assuming a more compact cloud than modelling C1 because the variability seen in the latter is more shallow, which at the same time explains why C1 lightcurve has a higher normalization despite the comparable distance from Sgr A*. No  adjustments in the flux normalizations were necessary, and no arbitrary normalization constants involved.   

We note that Sgr B2 is the most massive complex among molecular clouds in the Galaxy:
its mass has been estimated of about $10^7M_{\odot}$ (Lis \& Goldsmith 1989).
This cloud is also quite large, with clumps of molecular gas around the dense core.  
Most of X-ray flux is seen as a diffuse component, whereas numerous unresolved 
point sources within the cloud volume provide much less 
flux in the total energy budget. Even if the cloud geometry is complicated
(non-spherical), Murakami et al. (2000) approximate H$_2$ distribution in the cloud
by a general smooth profile of the form 
$n_{{\rm{}H}_2}\approx5.5\times10^4R_{1.25}^{-2}+2.2\times10^3$, where
$n_{{\rm{}H}_2}$ is the number density per cubic centimetre, while $R_{1.25}$ is 
radius in units of $1.25$~pc. 

Given the high central density and the overall size
of this system, the hydrogen absorption column density along the line of sight is estimated to be
around ($5$--$10)\times 10^{23}\;{\rm{}cm}^{-2}$. The dominant contribution to the hydrogen column 
should originate within the molecular cloud itself, which brings it to 
the verge of being a Compton thick medium, whereas other GC molecular clouds
are commonly thought not to be Compton thick. The optical thickness of the cloud
has implications for the expected shape of the reflection spectrum (Odaka et al. 2011), and it can increase the reprocessing timescale. Such models can only be calculated using a complex Monte Carlo numerical method to follow the time evolution of the reprocessed radiation, which is generally too slow, i.e., computationally far too expensive. The currently available quality of the data does not justify such an effort while modelling the time-dependent delays. Moreover, in this case the result depends on the exact geometry.

Our model points toward a possible origin for the Sgr A* activity. A sequence of clumps fall onto the black hole, all of them with similar values of mass and angular momentum. This suggests a single origin of the accreted clumps. A body -- cloud or star -- has probably been disrupted, and so it formed a stream of smaller gaseous bodies following roughly the same trajectory (like in a reminiscence of the famous Shoemaker-Levy comet impacting onto Jupiter in 1994).  The total mass accreting in the process (counting from 1400 to 1930; the earlier activity of Sgr~A* is completely unconstrained) is $\sim0.1$--$0.2 M_{\odot}$. All the clumps must have a common origin, and the disruption of the parent body must have happened a long time ago, since a period of time was needed for a blob segregation to grow owing to various gravitational perturbations. Segregation due to the interaction with surrounding gas would produce a more complicated profile of the mass function, which apparently has not happened. 

We would like to point out that within the frame of the model we can reproduce the reflection from the clouds B2 and C1, although the luminosity drop in the past activity period to the present level of quiescence is reproduced only partially. The discrepancy level depends to some extent on the data set used. In fact, there is an apparent offset between the flux levels derived from Ryu et al. (2013) versus Capelli et al. (2012), which can be attributed mainly to the uncertainties in the 3D positions of the reflection clouds and the difficulty of measuring column densities of X-ray nebulae. However, in both cases the final decay is more rapid than in the model if we consider a reasonable value of viscous timescales of 3 yr or above. In this paper we assumed that the rapid decay in the observed lightcurve is real, and therefore we constrained the viscous timescales to the short values when modelling the reflection from the molecular clouds as well. However, relaxing this constraint is possible in our scheme. If we remove the assumption about the recent rapid decay in Sgr~A* activity, we can fit the time evolution of the B2 reflection with parameters $t_{\rm visc} = 9$~yr, $R_{\rm cloud} = 3.5$~pc even in the constant density scheme.

The apparent problem of the too rapid decay in the lightcurve can be solved if we assume that the accretion rate in any single event has a cut-off at some level owing to interaction with the ambient medium. For example, thermal conduction can efficiently lead to the disappearance of tiny remnants of the cloud (Cowie \& McKee 1977). Such an interaction is indeed seen in numerical results; however, it does not seem to lead automatically to such a sharp cut-off (Guillochon \& Ramirez-Ruiz 2013).

\section {Conclusions}
\label{sec:conclusions}
X-rays from Sgr A* surrounding molecular clouds  suggest that the central black hole was temporarily more active in a historically recent past than it is at present (Sunyaev et al. 1993; Koyama et al. 1996; Sunyaev \& Churazov 1998). It is possible that Sgr A* reached the characteristic level for low-luminosity active galactic nuclei that can be detected in reflection by the clouds (Ponti et al. 2010; Terrier et al. 2010). Although the evidence of this bright period is still circumstantial, it is interesting to check whether there is enough material and suitable conditions in the region that could provide both the matter and energy source for the black hole on the right timescale and the phase appropriate to trigger and sustain the accretion over several hundred years. 

The mini-spiral of the Galactic centre is a potential source of material. This puzzling feature, located at a distance about $0.1$--$0.2$~pc (projected distance $\sim 0.06$~pc) from the supermassive black hole has been recently explained as consisting of three independent clumpy streams of mainly gaseous material at roughly Keplerian motion around the centre. The streams collide at some point (Zhao et al. 2010; Kunneriath et al. 2012), and this collision may cause the loss of the angular momentum and occasional inflow of clumpy material towards the black hole. A more detailed modelling of the accretion stream and the systematic stepping over the parameter space are needed to constrain the suitable distribution of the angular momentum of the clouds arriving to the black hole vicinity from greater distances of the mini-spiral region (Czerny et al. 2013). Numerical simulations of multiple colliding streams were recently performed in Alig et al. (2013), and these results can be relevant also in the context of the mini-spiral formation.

Despite the simplicity of this scheme, it does capture the basic smearing effect that acts 
on the intrinsic flare over the light-crossing time of the cloud. Scattering on the small cloud produces an observed lightcurve profile closely 
resembling the intrinsic flare, while a large one enhances the smearing 
of the lightcurve, and the decay part becomes symmetric with respect to the onset part. This 
general tendency is expected, but it is interesting to notice that the cloud size can be as large
as $\sim5$--6 pc and still produce a good formal fit to the data points (in terms of 
$\chi^2_{\rm{}red}\simeq1.1$ statistics). This size agrees with the expectation for the size of the core 
of Sgr B2. A still larger size for the cloud, however, would smear the decay part of the lightcurve too much, and 
then the formal fit also goes bad. The timescale of the flare decline can be about nine years, assuming that 
the lightcurve decline proceeds more closely according to Capelli et al. (2012), i.e. more slowly than the relatively rapid
drop seen in Ryu et al. (2013). Further monitoring of the B2 and other clouds is therefore needed.

As an illustration of the model we considered the fate of the G2 cloud that has been reported as falling towards Sgr~A*. As the cloud is disrupted, some material should settle near the pericentre radius. The viscous timescale corresponding to this radius is roughly 18~yr, assuming efficient heating and formation of an optically thin torus (ADAF). The peak luminosity due to the accretion onto Sgr A* should happen in $\sim 2022$ (under the assumptions of inefficient cooling). The mass of the cloud is quite low, so a significant enhancement (by a factor of ten) of the Sgr~A* luminosity is only expected if the viscous timescale of the inflow is this short. On the other hand, the increase in the source activity is expected to be much less spectacular if the timescale is longer, and in addition a considerable fraction of the material does not reach the black hole. Of course, before this happens we are likely to detect the emission due to shocks connected with the settlement of the gas onto the circularization radius, or if the cloud is not disrupted at all, we will observe the synchrotron emission from the bow shock (Narayan et al. 2012). The excess emission about $10$ Jy at frequencies $\sim1$--$10$~GHz can be expected, well above the quiescent level of emission from Sgr~A∗. Quite a different scenario would be expected if a star resides inside the cloud core (Eckart et al. 2013a,b).

In addition to the opportune case of the upcoming passage of the G2 cloud through the orbit pericentre, we suggest that the model of multiple accretion events can be relevant and useful in the context of other supermassive black holes in galactic nuclei, in particular for active galactic nuclei that are thought to be embedded by clumpy material with a relatively large filling factor. Very likely, the accretion episodes, as envisaged by this scheme, should occur frequently, and they should partly contribute to the notorious activity and the resulting variability of these sources.

\begin{acknowledgements}
The research leading to these results has received funding from Polish grant NN 203 380136, the COST Action ``Black Holes in a Violent Universe'' (ref. 208092), and the European Union Seventh Framework Programme (FP7/2007--2013) under grant agreement No.\ 312789. DK and VK acknowledge support from the collaboration project between the Czech Science Foundation and Deutsche Forschungsgemeinschaft (GACR-DFG 13-00070J). The Astronomical Institute is operated under the program RVO:67985815.
\end{acknowledgements}

\end{document}